# Analytical results for a Bessel function times Legendre polynomials class integrals

A A R Neves, L A Padilha, A Fontes, E Rodriguez, C H B Cruz, L C Barbosa and C L Cesar

Instituto de Física Gleb Wataghin, Universidade Estadual de Campinas, P.O. Box 6165, 13083-970, Campinas, Brazil.

E-mail: aneves@ifi.unicamp.br

**Abstract**. When treating problems of vector diffraction in electromagnetic theory, the evaluation of the integral involving Bessel and associated Legendre functions is necessary. Here we present the analytical result for this integral that will make unnecessary numerical quadrature techniques or localized approximations. The solution is presented using the properties of the Bessel and associated Legendre functions.

## 1. Introduction

On a recent work to calculate the optical force of the optical tweezers in a complete electromagnetic treatment for any beam shape focused at an arbitrary position we encountered an integral involving Bessel and associated Legendre functions [1]. The same integral appears in fields related to vector diffraction theory were computationally intensive methods or approximations are employed [2]. This article presents the analytical evaluation of this integral (1).

Typically, the first task in solving scattering problems is to decompose the incident beam into partial waves involving associated Legendre polynomials $P_n^m(\cos\theta)$ for the angular part and spherical Bessel functions $j_n(kr)$ for the radial part. The difficulty in doing this exactly, is to determine an analytical expression for integrals of the class,

$$I_n^m = \int_0^\pi d\theta \sin\theta \exp(iR\cos\alpha\cos\theta) P_n^m(\cos\theta) J_m(R\sin\alpha\sin\theta) \qquad (1)$$

which, so far, have not been reported in a closed form, as far as we know. This integral is not shown in any Integral Tables, nor in calculation packages such as Mathematica, and we do not know of any other report of this result. An analytical expression for this integral would be useful for many different fields in physics, especially for those that require partial wave decomposition. Without this result people have used all sorts of approximations to proceed forward and obtain results [3].

## 2. General Analysis

We analyzed the behavior of such integrals for the limit of $kr = R \to 0$ using Lock's result for the integral [4]:

$$\int_0^\pi d\theta \sin^{|m|+1}\theta \exp(\pm iR\cos\theta) P_n^{|m|}(\cos\theta) = 2(\pm i)^{n+|m|} \frac{(n+|m|)!}{(n-|m|)!} \frac{j_n(R)}{R^{|m|}} \quad (2)$$

and found that the lowest order term would be compatible with the closed form and simple expression given by:

$$I_n^m = 2i^{n-m} P_n^m(\cos\alpha) j_n(R) \quad (3)$$

valid for any $n \geq 0$ and $-n \leq m \leq n$. We then tested it numerically using Mathematica software (version 5.2) for randomly generated $n$, $m$, $\alpha$ and $R$ and showed that this result is indeed true. It would also allow one to access a whole family of integrals by taking any number of derivatives respect to $R$ or $\alpha$, such as:

$$\int_0^\pi d\theta \sin\theta \frac{d}{dR}\left[\exp(iR\cos\alpha\cos\theta) J_m(R\sin\alpha\sin\theta)\right] P_n^m(\cos\theta) = 2i^{n-m} P_n^m(\cos\alpha) j_n'(R) \quad (4)$$

After discovering the expression for this integral we proceed to prove it using the same induction procedure used in the Lock integral paper. We started by proving that $I_n^m$ follows the same recurrence relations as $F_n^m = 2i^{n-m} P_n^m(\cos\alpha) j_n(R)$. We then proved that the results holds for, $n = m = 0$.

The recurrence relations for the $I_n^m$ can be obtained by the associated Legendre polynomials recurrence relations:

$$\begin{aligned}(2n+1)\sin\theta P_n^m(\cos\theta) &= P_{n-1}^{m+1}(\cos\theta) - P_{n+1}^{m+1}(\cos\theta) \\ &= (n-m+1)(n-m+2)P_{n+1}^{m-1}(\cos\theta) - (n+m)(n+m-1)P_{n-1}^{m-1}(\cos\theta)\end{aligned} \quad (5)$$

where we adopted the sign convention for the associated Legendre polynomials of Abramowitz and Stegun [5], followed by the Mathematica software. The Bessel functions recurrence relation:

$$J_m(R\sin\alpha\sin\theta) = \frac{R\sin\alpha\sin\theta}{2m}\left[J_{m-1}(R\sin\alpha\sin\theta) + J_{m+1}(R\sin\alpha\sin\theta)\right] \quad (6)$$

Using relations (5) and (6) it can be readily shown that:

$$I_n^m = \frac{R\sin\alpha}{2m(2n+1)}\left[(n-m+1)(n-m+2)I_{n+1}^{m-1} - (n+m)(n+m-1)I_{n-1}^{m-1} + I_{n-1}^{m+1} - I_{n+1}^{m+1}\right] \quad (7)$$

is the desired recurrence relation for the $I_n^m$.

Now, for $F_n^m = 2i^{n-m} P_n^m(\cos\alpha) j_n(R)$ we used the associated Legendre polynomials recurrence relation:

$$\begin{aligned}\frac{2m}{\sin\alpha} P_n^m(\cos\theta) &= -(n-m+1)(n-m+2)P_{n+1}^{m-1}(\cos\alpha) - P_{n+1}^{m+1}(\cos\alpha) \\ &= -(n+m)(n+m-1)P_{n-1}^{m-1}(\cos\alpha) - P_{n-1}^{m+1}(\cos\alpha)\end{aligned} \quad (8)$$

and the spherical Bessel Functions relation:

$$j_n(R) = \frac{R}{2n+1}\left[j_{n-1}(R) + j_{n+1}(R)\right] \quad (9)$$

to prove that:

$$F_n^m = \frac{R\sin\alpha}{2m(2n+1)}\left[(n-m+1)(n-m+2)F_{n+1}^{m-1} - (n+m)(n+m-1)F_{n-1}^{m-1} + F_{n-1}^{m+1} - F_{n-1}^{m+1}\right] \quad (10)$$

This assures that both sides of the identity follow the same recurrence relation. Therefore, the only task remaining is to prove that:

$$I_0^0 = \int_0^\pi d\theta \sin\theta \exp(iR\cos\alpha\cos\theta) J_0(R\sin\alpha\sin\theta) = 2 j_0(R) \quad (11)$$

holds for $n = m = 0$. Now, the series expression for $J_0$ is,

$$J_0(R\sin\alpha\sin\theta) = \sum_{s=0}^{\infty}\frac{1}{2^{2s}}\frac{(-1)^s}{s!s!}R^{2s}(\sin\alpha)^{2s}(\sin\theta)^{2s} \tag{12}$$

therefore:

$$I_0^0(R,\cos\alpha) = \sum_{s=0}^{\infty}\frac{1}{2^{2s}}\frac{(-1)^s}{s!s!}R^{2s}(\sin\alpha)^{2s}\int_0^{\pi}d\theta\sin\theta\exp(iR\cos\alpha\cos\theta)(\sin\theta)^{2s} \tag{13}$$

The Poisson integral representation of the spherical Bessel function is:

$$j_s(R) = \frac{R^s}{2^{s+1}s!}\int_0^{\pi}d\theta\sin\theta\cos(R\cos\theta)(\sin\theta)^{2s} \tag{14}$$

which, added to the parity null term

$$\int_0^{\pi}d\theta\sin\theta\sin(R\cos\theta)(\sin\theta)^{2s} = 0 \tag{15}$$

can be rewritten as

$$\int_0^{\pi}d\theta\sin\theta\exp(iR\cos\alpha\cos\theta)(\sin\theta)^{2s} = \frac{2^{s+1}s!}{R^s}j_s(R) \tag{16}$$

This leaves the initial integral as:

$$I_0^0(R,\cos\alpha) = \sum_{s=0}^{\infty}\frac{1}{2^s}\frac{(-1)^s}{s!}R^s(\sin\alpha)^{2s}\frac{2}{(\cos\alpha)^s}j_s(R\cos\alpha) \tag{17}$$

That is still not good because the argument of the spherical Bessel function is $R\cos\alpha$ instead of $R$. To change the argument we use the multiplication theorem for Bessel functions:

$$J_\nu(\lambda z) = \lambda^\nu\sum_{s=0}^{\infty}\frac{1}{2^s}\frac{(1-\lambda^2)^s}{s!}z^s J_{\nu+s}(z) \tag{18}$$

To obtain,

$$j_n(z) = \sqrt{\frac{\pi}{2z}}J_{n+1/2}(z) \tag{19}$$

we can now make $\nu = 1/2$, $n=0$, $z = R\cos\alpha$ and $\lambda = 1/\cos\alpha$. Therefore:

$$j_0(R) = \frac{1}{\sqrt{\cos\alpha}}\sum_{s=0}^{\infty}\frac{(-1)^s}{2^s}\frac{(1-\cos^2\alpha)^s}{s!(\cos\alpha)^{2s}}(R\cos\alpha)^s\sqrt{\frac{\pi}{2R}}J_{s+1/2}(R\cos\alpha) \tag{20}$$

and finally

$$j_0(R) = \sum_{s=0}^{\infty}\frac{(-1)^s}{2^s}\frac{(\sin\alpha)^{2s}}{s!(\cos\alpha)^s}R^s j_s(R\cos\alpha) \tag{21}$$

Comparing this result with the series obtained it turns out that

$$I_0^0 = 2i^0 P_0^0(\cos\alpha)j_0(R) \tag{22}$$

and the proof is complete, validating the following integral for all $n$ and $m$.

$$\int_0^{\pi}d\theta\sin\theta\exp(iR\cos\alpha\cos\theta)P_n^m(\cos\theta)J_m(R\sin\alpha\sin\theta) = 2i^{n-m}P_n^m(\cos\alpha)j_n(R) \tag{23}$$

### 3. Concluding remarks

In this paper, we have shown the analytical solution to an integral involving Bessel and associated Legendre functions (23). We have obtained the solution through recurrence relations and the multiplication theorem. We believe that this result, analytical and simple is of interest to the general community especially for problems involving electromagnetic vector diffraction of arbitrary beams.

**Acknowledgements**

This work was partially supported by Fundação de Amparo à Pesquisa do Estado de São Paulo (FAPESP) through the Optics and Photonics Research Center (CePOF). We thank the Coordenação de Aperfeiçoamento de Pessoal de Nível Superior (CAPES) for financial support of this research.